\documentclass[10pt,letterpaper]{article}
\usepackage[top=0.85in,left=2.75in,footskip=0.75in,marginparwidth=2in]{geometry}

\usepackage[utf8]{inputenc}

\usepackage{cite}

\usepackage{nameref,hyperref}
\usepackage[numbers]{natbib}

\usepackage{microtype}
\DisableLigatures[f]{encoding = *, family = * }

\raggedright
\usepackage{changepage}

\usepackage[aboveskip=1pt,labelfont=bf,labelsep=period,singlelinecheck=off]{caption}

\makeatletter
\renewcommand{\@biblabel}[1]{\quad#1.}
\makeatother

\usepackage{lastpage,fancyhdr,graphicx}
\usepackage{epstopdf}
\fancyheadoffset[L]{2.25in}
\fancyfootoffset[L]{2.25in}

\usepackage{color}

\definecolor{Gray}{gray}{.25}

\usepackage{graphicx}

\usepackage{sidecap}
\usepackage{caption}
\usepackage{subcaption}

\usepackage{wrapfig}
\usepackage[pscoord]{eso-pic}
\usepackage[fulladjust]{marginnote}
\reversemarginpar
\usepackage{times}
\usepackage{multirow}

\usepackage{adjustbox}
\usepackage{longtable}
\usepackage{rotating}

\usepackage[table]{xcolor}

\begin{document}
\vspace*{0.35in}

\begin{flushleft}
{\Large
\textbf\newline{Student gender modulates the intersection of calculus proficiency and calculus self-efficacy in an introductory electricity and magnetism course}
}
\newline
\\
Christopher J. Fischer\textsuperscript{*},
Jennifer Delgado,
Sarah LeGresley,
Jessy Changstrom
\\
\bigskip
Department of Physics and Astronomy, University of Kansas, 1251 Wescoe Hall Drive, Lawrence, KS, 66049
\\
\bigskip
* shark@ku.edu

\end{flushleft}

\section*{Abstract}
    We assessed changes in calculus proficiency and calculus self-efficacy in a second semester course of introductory physics focused on electricity and magnetism.  While all students demonstrated an increase in calculus proficiency, including a possible improvement in calculus transfer to physics, women displayed larger gains than men.  Conversely, men showed larger gains in calculus self-efficacy.  When combined, these data suggest that student identity modulates the correlation between a student’s calculus abilities and their perception or self-evaluation of those abilities.   These data highlight a potential contributing factor to gender-related differences in physics self-efficacy as well as the complexity of addressing those differences.


\section{Introduction}

The ability to use mathematics is an essential skill common to all science and engineering disciplines.  Because of this and due to economic and logistic constraints, courses in introductory mathematics at most colleges and universities are shared by many degree programs and centrally taught in a department of mathematics.  Unfortunately, this structure can result in the mathematics being decontextualized, thereby placing on students the burden of transferring their skills in solving mathematical problems to other STEM areas.  However, while students are generally expected to be capable of this transference, evidence suggests that it actually poses a large barrier to student learning in physics \cite{britton2002students,cui2006assessing,Rebello2007,new2012researching}.  Indeed, the emphasis on applied mathematics in introductory physics courses is believed to be responsible for the observed positive correlation between mathematics ability and student performance in these courses \cite{hudson1977correlation,cohen1978cognitive,hudson1981correlation,brekke1994some}.  
Although we and others have suggested how calculus proficiency in introductory physics and possibly mathematics transfer to physics may be improved  \cite{PhysRevPhysEducRes.15.020126,kezerashvili2007transfer,wagner2012representation,turcsucu2017teachers}, a quantitative analysis of how curriculum changes and/or interventions in introductory physics affect student ability to solve calculus problems in the context of physics has not yet been completed.  As part of an ongoing effort to address this need, we report here our analysis of proficiency with integral calculus in the context of introductory physics, covering electricity and magnetism, combined with an independent assessment of calculus self-efficacy.  We show that while a positive correlation between student self-assessment of calculus ability and student proficiency with calculus exists for men, a negative correlation between these traits is present for women, even though women perform equal to or better than men on assessments of calculus proficiency.  These results thus build off our previous work indicating an identity-modulated  correlation between physics self-efficacy and calculus proficiency \cite{fischer2021interplay}.

\section{Background}

The suggested schedule for most calculus-based STEM (\textit{e.g.}, physical sciences and engineering) degree programs at the University of Kansas has students enrolling in General Physics II (the second semester of calculus-based introductory physics covering electricity and magnetism) during their third semester at the same time they are enrolled in the third semester of calculus; General Physics I is taken alongside the second semester of calculus.  The coordinating scheduling of calculus and introductory physics courses allows us to task students with using calculus continually in their physics courses \cite{PhysRevPhysEducRes.15.020126,Fischer2019a,Fischer2019b} and thus to assess how well students transfer their calculus knowledge and skills to physics.  When characterizing and analyzing this knowledge transfer, we use the model proposed by Rebello \textit{et al.} \cite{Rebello2007}, which builds upon earlier work by Gagn\'{e} \cite{gagne308conditions} and Reddish \cite{redish2004theoretical}.  In this context, we make a distinction between horizontal transfer, which describes the use of new information in existing mental models and vertical transfer in which an existing mental model does not exist, thus requiring the creation of new mental models \cite{cui2006assessing,fischer2021interplay}.

Self-efficacy is a measure of one's belief to succeed \cite{sawtelle2012exploring,nissen2016gender,espinosa2019reducing}.  As such, it can affect many dimensions of learning including motivation \cite{prat2010interplay}, interest \cite{zimmerman2000self}, engagement \cite{schunk2002development}, and how difficult tasks are perceived \cite{watt2006role}.  Not surprisingly, self-efficacy influences both performance and retention in STEM courses \cite{sawtelle2012exploring} and women often have lower self-efficacy in physics than men despite performing similarly or better than men in their physics courses \cite{nissen2016gender,marshman2018female,espinosa2019reducing}. 

\subsection{Research Questions}  This work seeks to address the following research questions (RQ) in the context of our calculus-based introductory physics courses:
\begin{quote}
    \textbf{RQ1.} How does the emphasis on calculus in introductory physics courses affect student ability with pure and applied calculus?
\end{quote}

\begin{quote}
    \textbf{RQ2.} How is student self-efficacy for using calculus in introductory physics correlated with student proficiency for pure and applied calculus?
\end{quote}

\section{Methodology}

Students in General Physics II at the University of Kansas are tasked with using integral calculus to calculate the electric and magnetic fields and potentials of distributions of electric charge, electric and magnetic flux, and the electric and magnetic forces and potential energies associated with the interactions of charged systems.  Our survey for assessing calculus proficiency in this course builds upon a previously implemented assessment \cite{fischer2021interplay} and consists of the following six questions:

\vspace{2mm}

\noindent \textbf{C1.} Complete the following integral $\int \limits_0^6 x^2 dx$

\vspace{1.5mm}

\noindent \textbf{C2.} Complete the following integral $\int \limits_0^1 \frac{2x}{\left( x^2 + 1 \right)^2} dx$

\vspace{1.5mm}

\noindent \textbf{C3.} The population of a town, denoted by the variable $P$, increases at a rate described by the following equation: $\frac{dP}{dt} = 20 + 0.6t^2$.  The variable $t$ is the time.  What is the population of the time at $t=5$ if the population at $t=0$ is 0.

\vspace{1.5mm}

\begin{wrapfigure}{r}{0.2\textwidth}
\label{fig:prelimmta2}
  \begin{center}
    \includegraphics[width=0.2\textwidth]{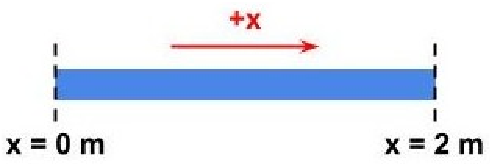}
  \end{center}
\end{wrapfigure}

\noindent \textbf{C4.} The electric potential energy $\left( U_e \right)$ of a point particle with electric charge $q$ in the presence of an external electric potential $\left( V_e \right)$ is $U_e = q V_e$.  The uniformly charged straight and rigid rod shown in the figure is in the presence of the following external electric potential: $V_e = \left( 4 \mathrm{\frac{J}{Cm^2}} \right)x^2$.  The variable $x$ in this equation denotes the position along the $x$-axis as defined in the figure.  The rod has a length of 2 m and a constant electric charge density of $3 \mathrm{\frac{C}{m}}$.  Determine the electric potential energy associated with the interaction between the rod and the external electric potential.

\vspace{1.5mm}

\begin{wrapfigure}{r}{0.2\textwidth}
\label{fig:prelimmta1}
    \includegraphics[width=0.2\textwidth]{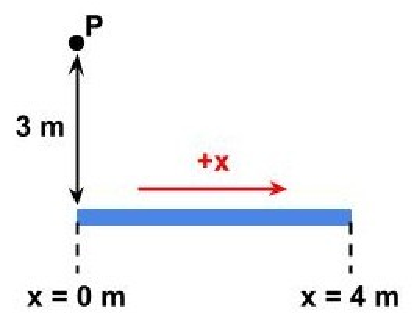}
      \label{fig:prelimmta}    
\end{wrapfigure}

\noindent \textbf{C5.} The electric potential $\left( V_e \right)$ a distance $r$ away from a point particle with electric charge $q$ is described by the following equation: $V_e = \frac{q}{4 \pi \epsilon r}$.  The straight and rigid rod shown in the figure to the right has a length of 4 m and a non-uniform linear electric charge density described by the following equation: $\left( 2 \pi \times 10^{-9} \mathrm{\frac{C}{m^2}}   \right) x$.  The variable $x$ in this equation denotes position along the $x$-axis as defined in the figure.  Determine the electric potential of this rod at the position $P$ that is a distance 3 m above one end of the rod as shown in the figure.  The variable $\epsilon = 1 \times 10^{-11} \mathrm{\frac{C}{Vm}}$.

\vspace{1.5mm}

\begin{wrapfigure}{r}{0.2\textwidth}
\label{fig:prelimmta2}
  \begin{center}
    \includegraphics[width=0.2\textwidth]{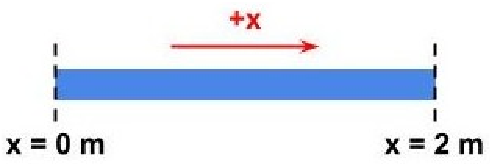}
  \end{center}
\end{wrapfigure}

\noindent \textbf{C6.}  The moment of inertia $\left( I \right)$ of a point particle with mass $m$ located a distance $x$ away from an axis of rotation is $I = mx^2$.  The straight and rigid rod shown in the figure to the right has a length of 2 m and non-uniform mass density described by the following equation: $\left( 5 \mathrm{\frac{kg}{m^3}}  \right) x^2$.  The variable $x$ in this equation denotes position along the $x$-axis as defined in the figure.  Determine the moment of inertia of this rod for rotations around an axis located at $x = 0 \mathrm{m}$.  

We denote questions C1, C2, and C3 on this survey as ``pure'' calculus questions.  Conversely, we denote questions C4, C5, and C6 as ``applied'' calculus questions as they require students to use calculus within a physics context.  We also note that the integration required to solve C4 and C6 is equivalent to the integration required to solve C1.  Similarly, the integration required to solve C5 is equivalent to the integration required to solve C2.  Finally, since students at our institution have not used calculus to determine moments of inertia in General Physics I or General Physics II \cite{fischer2021interplay}, solving C6 requires represents a new application of integral calculus in the context of physics.  However, by prompting students with the equation for the moment of inertia of a point particle, we hoped that student would recognize that the integration necessary for the solution is similar to the integration necessary to solve C1 and C4.

\begin{figure}[h]
\begin{center}
\includegraphics[width=0.7\textwidth]{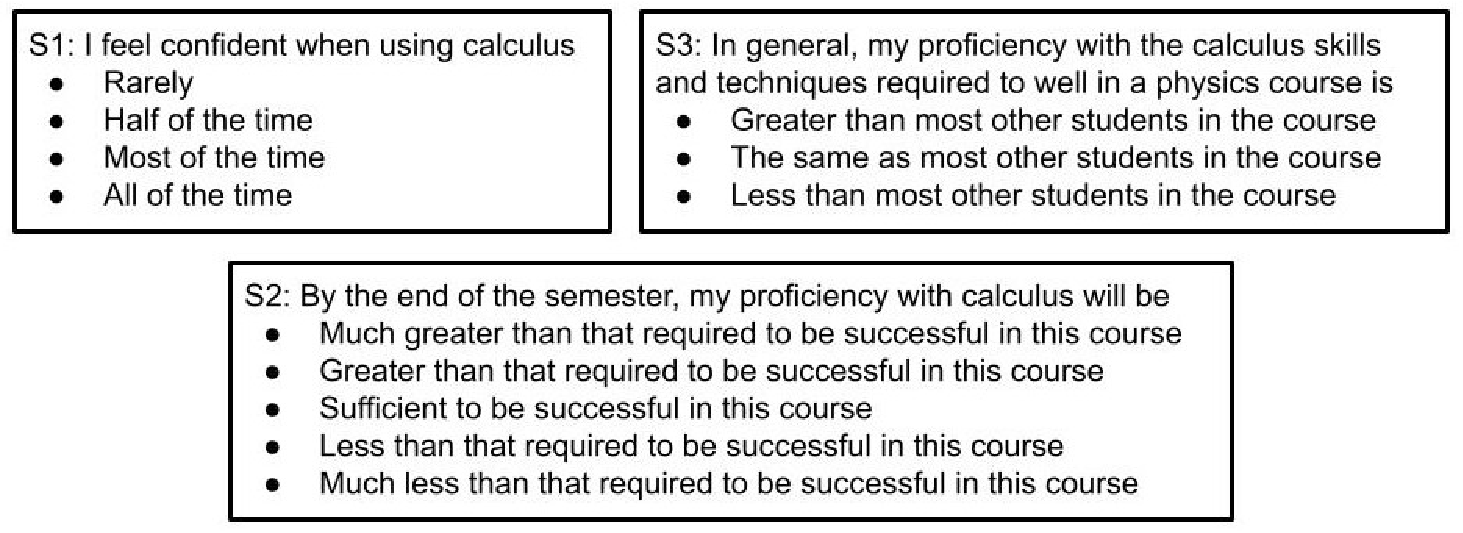}
\end{center}
\caption{The calculus self-efficacy questions used in this study.}
\label{fig:mathefficacy}
\end{figure}

Our assessment of calculus self-efficacy \cite{fischer2021interplay} consists of the three questions shown in Figure \ref{fig:mathefficacy}.  The calculus self-efficacy survey and calculus proficiency survey were administered at the beginning (survey 1), in the middle (survey 2), and at the end (survey 3) of the semester in General Physics II during the Summer 2021 and Fall 2021 semesters.  The responses to survey questions were graded as favorable, neutral, or unfavorable, and given a corresponding numeric score of 3, 2, or 1; for the calculus proficiency survey, a correct answer was scored favorable and an incorrect answer was scored unfavorable.  

The total enrollment in the courses where the surveys were deployed was 362 students, however, not all students consented to their survey results being included in the analysis presented here.  We further restricted our data set to only those students who completed all self-efficacy surveys and all calculus proficiency surveys so that our analysis was conducted on matched students only.  This leaves a total of 293 students in General Physics II (205 men and 88 women) included in the results presented here.  Please note that since we are relying upon institutional data for creating these cohorts of students, these divisions do not therefore necessarily reflect how students identify personally.

\section{Results}

\subsection{Validation of the Survey}

\begin{figure}[h]
\begin{center}
\includegraphics[width=0.5\textwidth]{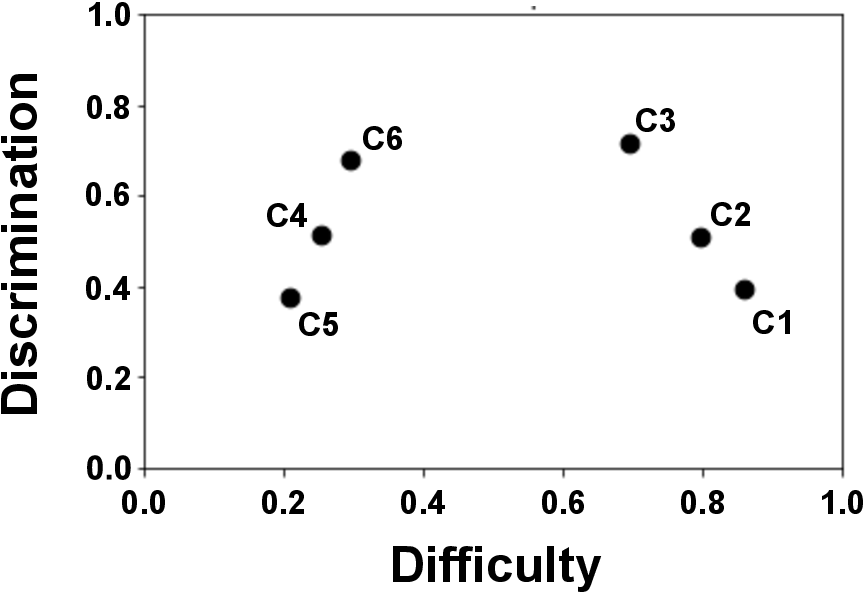}
\end{center}
\caption{The discrimination and difficulty of the ``pure'' and ``applied'' calculus questions.}
\label{fig:discrimination}
\end{figure}

In our first analysis of the survey results, we calculated the discrimination and difficulty of each calculus proficiency question \cite{baker2001basics,salkind2017tests}; note that difficulty is defined as the fraction of students answering the question correctly \cite{baker2001basics,salkind2017tests}.  As shown in Figure \ref{fig:discrimination}, a smaller fraction of students answered the ``applied'' calculus questions (C4, C5, and C6) correctly than answered the ``pure'' calculus questions (C1, C2, and C3) correctly.  This is consistent with the expectation that ``applied'' calculus questions would be more difficult for students, perhaps owing to difficulties of using calculus in a physics context (\textit{i.e.}, with calculus transfer to physics).  All questions showed a discrimination above 0.35, suggesting that these questions provide moderate to high discrimination \cite{salkind2017tests}.

To establish the separability of three different categories of questions on the assessment - pure calculus questions, applied calculus questions, and calculus self-efficacy questions - we performed an exploratory factor analysis on the items in the survey.  A principal components analysis with Varimax
rotation method was used, and the initial eigenvalues indicated that the first three components (all with eigenvalues greater than 1) explained a total of 51\% of the variance.  The factor loadings determined from these principal components are shown in Table \ref{tab:factor_loadings}.  Factor 1 is associated with calculus self-efficacy questions (S1, S2, and S3), factor 2 is associated with the ``pure calculus'' questions (C1, C2, and C3), and factor 3 is associated with the ``applied calculus'' questions (C4, C5, and C6).  Thus, the data supported the existence of three separate scales and items loaded on the scales as intended.  

When the fourth principal component is included in the analysis, the total variance explained increases to 62\%.  Factor 1 and factor 2 generated by this analysis are identical in composition to those determined using only the first three eigenvectors.  The newly determined factor 3 is associated with question C3 and question C6, and the new factor 4 is associated with question C5.  We ascribe this splitting into two factors to describe the ``applied calculus'' questions with the fact question C5 had the lowest correct response rate of any of the questions (Figure \ref{fig:discrimination}).  Regardless, this analysis is also consistent with the presence of three separate scales - pure calculus questions, applied calculus questions, and calculus self-efficacy questions - existing within the survey. 

\begin{sidewaystable}
   \caption{Factor loadings for each question on the assessment.  Loadings greater than 0.2 are  highlighted in gray.\label{tab:factor_loadings}}
    \begin{tabular}{c|c c c c c c c c c}
         \multirow{2}{*}{Factor} & \multicolumn{9}{c}{Question} \\
          & C1 & C2 & C3 & C4 & C5 & C6 & S1 & S2 & S3 \\
         \hline
         \hline
         F1 (calculus self-efficacy) &  -0.052 & 0.003 & -0.037 & -0.003 & 0.042 & -0.025 & \cellcolor{gray!25}0.416 & \cellcolor{gray!25}0.451 & \cellcolor{gray!25}0.462 \\
         F2 (pure calculus proficiency) & \cellcolor{gray!25}0.513 & \cellcolor{gray!25}0.466 & \cellcolor{gray!25}0.405 & -0.114 & -0.103 & 0.066 & -0.013 & -0.026 & -0.065\\
         F3 (applied calculus proficiency) &  -0.102 & -0.070 & 0.123 & \cellcolor{gray!25}0.514 & \cellcolor{gray!25}0.408 & \cellcolor{gray!25}0.524 & -0.112 & 0.01 & 0.051 \\
    \end{tabular}   
\end{sidewaystable}

\subsection{Survey Results}

\begin{figure}[h]
\begin{center}
\includegraphics[width=0.8\textwidth]{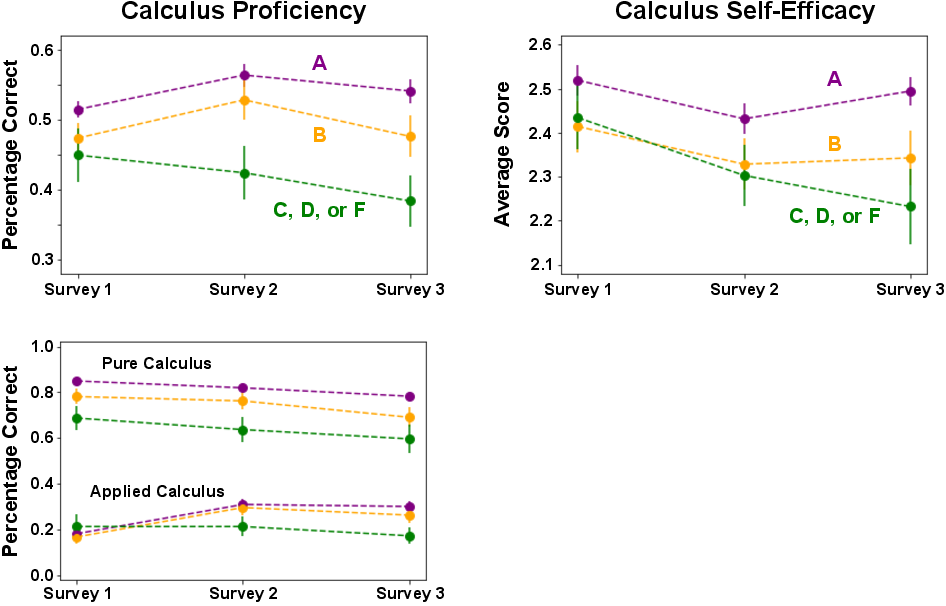}
\end{center}
\caption{Responses to calculus proficiency survey (left panels) and calculus self-efficacy survey (right panel) disaggregated by course grade.  190 students earned an A (purple), 70 students earned a B (orange), and 33 students earned a C, D, or F (green).  Error bars denote the 68\% confidence interval of the standard error.}
\label{fig:allresults}
\end{figure}

The results for the calculus proficiency survey and calculus self-efficacy survey are shown in Figure \ref{fig:allresults}.  In this figure, the average self-efficacy score is calculated by averaging the scores on the three calculus self-efficacy questions (Figure \ref{fig:mathefficacy}).  On average, students scoring higher on the calculus proficiency survey earned higher grades in the course, consistent with previous work showing a correlation between mathematics ability and student performance in physics courses \cite{hudson1977correlation,cohen1978cognitive,hudson1981correlation,brekke1994some}.  In addition, these data also indicate that students score higher on the ``pure'' calculus questions than on the ``applied'' calculus questions.  While this result may not be surprising for the first survey, and perhaps the second survey, as the students are still in the process of learning the associated physics content, it is nevertheless noteworthy that this trend persists on the third survey taken at the end of the semester.

\begin{figure}[h]
\begin{center}
\includegraphics[width=0.8\textwidth]{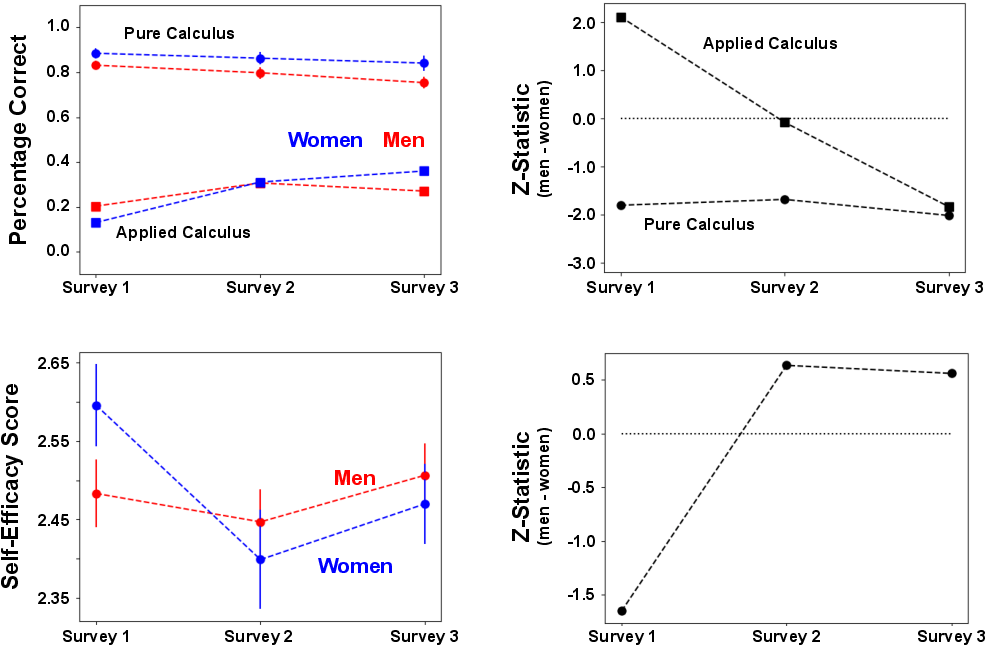}
\end{center}
\caption{Responses to surveys for women (blue) and men (red) earning an A in General Physics II.  The top panels are the percentage of correct answers for ``pure'' calculus (circles) and ``applied'' calculus questions (squares).  The bottom panels are the average score on the calculus self-efficacy questions.  All uncertainties are the 68\% confidence interval of the standard error.  The corresponding Z-statistic for the difference between the distributions for men and women are shown in the right panels.  All uncertainties are the 68\% confidence interval of the standard error.}
\label{fig:scores_A}
\end{figure}

As the data shown in Figure \ref{fig:allresults} demonstrate that scores on the calculus proficiency survey and calculus self-efficacy survey were dependent on course grade, we next sought to determine how student identity (\textit{i.e.}, student gender) modulates the interaction between calculus proficiency and calculus self-efficacy by focusing on students earning the same course grade.  The average responses to the surveys by students earning a grade of A in General Physics II (129 men and 61 women) are shown in Figure \ref{fig:scores_A}.  These data show that women score higher than men on the ``pure'' calculus questions on the calculus proficiency survey, with differences between 1 and 2 pooled standard errors, and that the difference between the scores of women and men increases throughout the semester.  Similarly, while men score higher than women on the ``applied'' calculus questions on the calculus proficiency survey at the beginning of the semester, by the end of the semester, women score higher than men on those questions too.  Figure \ref{fig:scores_A} also shows the differences between the average calculus self-efficacy scores for men and women earning an A in General Physics II.  While the average calculus self-efficacy score is higher for women than for men at the beginning of the semester, by the end of the semester, the situation has reversed.  When the data for calculus self-efficacy and calculus proficiency are combined, we see that as the semester progresses women score ever better than men on calculus proficiency  while scoring ever worse than men on the calculus self-efficacy.  Thus, these data indicate an inverse relationship between the perception of calculus proficiency and the demonstration of calculus proficiency.   

It is interesting to note that while the scores on the ``applied'' calculus questions increase throughout the semester, the scores on the ``pure'' calculus questions decrease.  Furthermore, as indicated by the Z-statistics shown in Figure \ref{fig:scores_A}, these changes are not of equal magnitude for men and women. 
 For the ``pure'' calculus questions, the average score for women decreased from $(89 \pm 2)\%$ to $(84 \pm 3)\%$, corresponding to a Z-statistic of -1.09, while the average score for men decreased from $(83 \pm 2)\%$ to $(75 \pm 3)\%$, corresponding to a Z-statistic of -2.31; the errors indicate the 68\% confidence interval for the standard error.  In contrast, for the ``applied'' calculus questions, the average score for women increased from $(13 \pm 2)\%$ to $(36 \pm 4)\%$, corresponding to a Z-statistic of 4.74, while the average score for men increased from $(20 \pm 2)\%$ to $(27 \pm 3)\%$, corresponding to a Z-statistic of 1.92; the errors indicate the 68\% confidence interval for the standard error.  Women thus displayed a larger improvement in scores on the ``applied'' calculus questions and a smaller decrease in scores on the ``pure'' calculus questions when compared to men.

\begin{figure}[h]
\begin{center}
\includegraphics[width=0.8\textwidth]{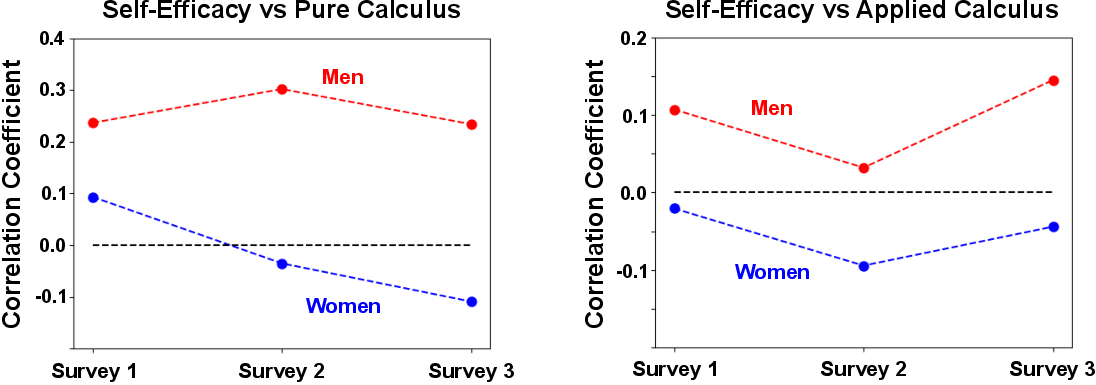}
\end{center}
\caption{Correlation coefficients between average self-efficacy scores, average ``pure'' calculus scores, and average ``applied'' calculus scores for men (red) and women (blue) earning an A in the course.}
\label{fig:correlation_A}
\end{figure}

To investigate this relationship further, we calculated the correlation between the average self-efficacy scores, average ``pure'' calculus scores, and average ``applied'' calculus scores for men and women, separately, for students earning an A in the course.  These correlation coefficients are shown in Figure \ref{fig:correlation_A} and indicate that the inverse relationship (\textit{i.e.}, negative correlation) between calculus proficiency and calculus self-efficacy exist within the population of women students, but not within the population of men students.  

\subsection{Principal Component Analysis}

\begin{figure}[h]
\begin{center}
\includegraphics[width=0.8\textwidth]{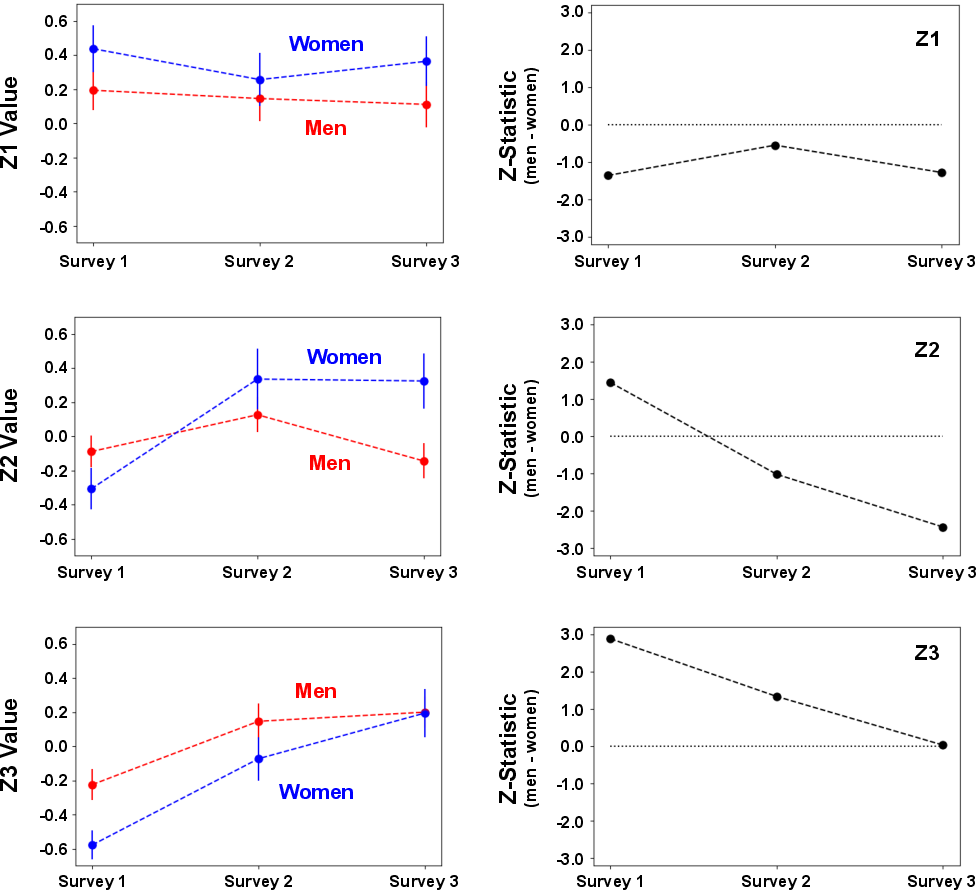}
\end{center}
\caption{The first (Z1), second (Z2), and third (Z3) principal components of the combined data set of pure math questions, applied math questions, and math self-efficacy questions for students earning an A in General Physics II.  The left panels show the mean and standard error (68\% confidence interval) of the values of these principal components for women (blue) and men (red).  The right panels show the corresponding Z-statistic for the difference between the values of the principal components for men and women.}
\label{fig:principal_components_212_A}
\end{figure}

Similar conclusions are found through a principal component analysis \cite{manly1994multivariate} of these data.  The coefficients of contributions of each survey question to the first three of these principal components and corresponding probability contributions are shown in Table \ref{tab:principalcomponents_212}; these principal components were calculated using all matched student responses (\textit{i.e.}, for all students, regardless of grade earned in the course).  The average values of the first three of these principal components for men and women earning an A in General Physics II are shown in Figure \ref{fig:principal_components_212_A}.  The first principal component, Z1, is dominated by the calculus self-efficacy questions and the ``pure'' calculus questions with the coefficients of all questions having the same sign (Table \ref{tab:principalcomponents_212}).  The values of Z1 are consistently higher for women than for men (Figure \ref{fig:principal_components_212_A}) as the difference in calculus self-efficacy between men and women is smaller than the difference in calculus proficiency between men and women.  In other words, the value of Z1 is consistently larger for women than for men since decreases in calculus self-efficacy for women occurring throughout the semester are compensated by concomitant smaller increases in calculus proficiency by men. 

As shown in Table \ref{tab:principalcomponents_212}, the coefficients for the contributions of the calculus questions to the second principal component, Z2, are all positive while the coefficients for the contributions from the calculus self-efficacy questions to Z2 are all negative.  Thus, the second principal component of these data represents the anti-correlation between student proficiency with calculus and their calculus self-efficacy.  As shown in Figure \ref{fig:principal_components_212_A}, the values of this principal component for men and women diverge over the course of the semester with the means of the populations separated by more than two pulled standard errors by the third survey.  In light of the results shown in Figure \ref{fig:scores_A}, we understand the origin of this divergence between men and women resulting from women scoring higher than men on the calculus proficiency survey and scoring less favorable than men on the calculus self-efficacy survey.

The third principal component, Z3, is dominated by questions from the calculus proficiency survey with ``applied'' calculus questions having positive coefficients and ``pure'' calculus questions having negative coefficients (Table \ref{tab:principalcomponents_212}).  Thus, a more positive value of Z3 corresponds to answering the ``applied'' calculus questions more favorably, the ``pure'' calculus questions less favorably, or some combination of both.  As shown in Figure \ref{fig:principal_components_212_A}, the values of Z3 are initially higher for men than for women, but the difference between men and women decreases steadily throughout the semester as the value of Z3 for women increases more than the value of Z3 for men.  These changes in Z3 thus reflect the fact that women are answering the ``pure'' calculus questions and the ``applied'' questions more favorably than men throughout the semester, with the gap between men and women increasing over the semester (Figure \ref{fig:scores_A}).    


\begin{sidewaystable}
\caption{The first three principal components for all students in General Physics II with negative coefficients highlighted in gray.  These principal components account for 21.7\% (Z1), 16.3\% (Z2), and 13.3\% (Z3) of the total variance in the data.  For Z2, the signs of the coefficients for calculus self-efficacy questions (S1, S2, and S3) are opposite the signs of the ``pure'' calculus questions (C1, C2, and C3) and the ``applied'' calculus questions (C4, C5, and C6).  For Z3, the signs of the coefficients for the ``pure'' calculus questions are opposite the signs for the ``applied'' calculus questions.\label{tab:principalcomponents_212}}
    \begin{adjustbox}{max width=\textwidth}
    \begin{tabular}{c | c | c | c || c | c | c | c || c | c | c | c}
    \multicolumn{4}{c}{\textbf{Z1}} & \multicolumn{4}{c}{\textbf{Z2}} & \multicolumn{4}{c}{\textbf{Z3}} \\
    \multirow{2}{*}{\textbf{Question}} & \multirow{2}{*}{\textbf{Coefficient}} & \multirow{2}{*}{\textbf{Probability}} & \textbf{Cumulative} & \multirow{2}{*}{\textbf{Question}} & \multirow{2}{*}{\textbf{Coefficient}} & \multirow{2}{*}{\textbf{Probability}} & \textbf{Cumulative} & \multirow{2}{*}{\textbf{Question}} & \multirow{2}{*}{\textbf{Coefficient}} & \multirow{2}{*}{\textbf{Probability}} & \textbf{Cumulative}\\
    & & & \textbf{Probability} & & & & \textbf{Probability} & & & & \textbf{Probability} \\
    \hline
    \hline
    S2 & 0.443 & 0.196 & 0.196 & C6 & 0.424 & 0.180 & 0.180 & C4 & 0.504 & 0.254 & 0.254 \\ 
    S3 & 0.438 & 0.192 & 0.388 & S1 & \cellcolor{gray!25}-0.416 & 0.173 & 0.352 & C1 & \cellcolor{gray!25}-0.446 & 0.199 & 0.454 \\
    C2 & 0.368 & 0.134 & 0.524 & S3 &  \cellcolor{gray!25}-0.370 & 0.137 & 0.489 & C5 & 0.419 & 0.175 & 0.629  \\
    S1 & 0.365 & 0.133 & 0.657 & S2 &  \cellcolor{gray!25}-0.366 & 0.134 & 0.623 & C6 & 0.386 & 0.149 & 0.778 \\
    C3 & 0.360 & 0.129 & 0.787 & C3 &  0.357 & 0.127 & 0.751 & C2 & \cellcolor{gray!25}-0.375 & 0.140 & 0.918\\
    C1 & 0.338 & 0.114 & 0.901 & C4 &  0.291 & 0.085 & 0.835 & S3 & 0.186 & 0.034 & 0.952 \\
    C6 & 0.258 & 0.066 & 0.968 & C1 & 0.278 & 0.077 & 0.913 & C3 & \cellcolor{gray!25}-0.181 & 0.033 & 0.985 \\
    C5 & 0.132 & 0.017 & 0.985 & C2 &  0.228 & 0.052 & 0.964 & S2 & 0.122 & 0.015 & 1.000\\
    C4 & 0.123 & 0.015 & 1.000 & C6 & 0.189 & 0.036 & 1.000 & S1 & 0.005 & $< 0.001$ & 1.000 \\
\end{tabular}
\end{adjustbox}
\end{sidewaystable}

\subsection{Network Analysis}

The inverse relationship between performance on the ``pure'' calculus questions and the ``applied'' calculus questions on the calculus proficiency survey is also clear from network analysis \cite{dalka2022network} of the combined results from the calculus proficiency survey and the calculus self-efficacy survey.  These networks were subsequently sparsified using previously published methods \cite{foti2011nonparametric,dalka2022network} and the NetworkX package \cite{SciPyProceedings_11} for Python \cite{van1995python} to elucidate the backbone of the corresponding networks.  We employed a significance level \cite{foti2011nonparametric,dalka2022network} of 0.3 for these sparisifications.

\begin{figure}[h]
\begin{center}
\includegraphics[width=0.8\textwidth]{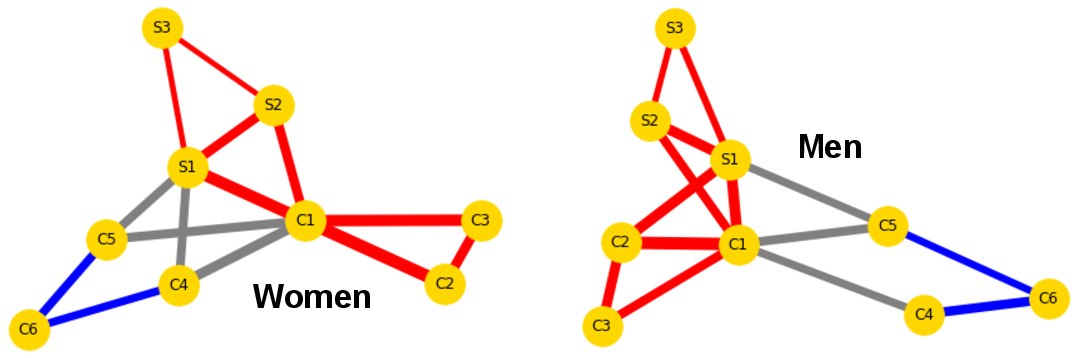}
\end{center}
\caption{The locally adaptive network sparsifications for students earning an A in General Physics II using a significance level of 0.3.  The questions on surveys are represented by the nodes of these graphs and the thickness of the edge denotes the weight of the connection \cite{dalka2022network}.  A red edge denotes a positive correlation between  responses to the connected questions, with a majority of the students answering both questions favorably.  A blue edge denotes a positive correlation between responses to the connected questions, with a majority of the students answering both questions unfavorably.  A grey edge denotes an inverse correlation between the responses to the questions.}
\label{fig:LANS_212_A}
\end{figure}

As shown in Figure \ref{fig:LANS_212_A}, the sparsified networks for all students earning an A in General Physics II show a positive correlation exists between responses to the ``applied'' calculus question (C4, C5, and C6) with the majority of students answering those questions incorrectly.  Responses to these questions are inversely correlated with responses to the ``pure'' calculus questions (C1, C2, and C3) and the calculus self-efficacy questions (S1, S3, and S3).  This inverse correlation between the ``pure'' calculus questions and the ``applied'' calculus questions is also manifest in the third principal component of the data for General Physics II (Table \ref{tab:principalcomponents_212}).  The modular communities \cite{newman2004finding}, determined using the greedy modularity maximization algorithm within the NetworkX package \cite{SciPyProceedings_11}, are the same for both networks.  The three communities within each network consist of the ``pure'' calculus questions (C1, C2, and C3), the ``applied'' calculus questions (C4, C5, and C6), and the calculus self-efficacy questions (S1, S2, and S3).  This result is consistent with the exploratory factor analysis discussed above and provides further support that three separate scales - pure calculus questions, applied calculus questions, and calculus self-efficacy questions - exist within the survey.

\section{Discussion}

The increase in correct responses to the ``applied'' calculus questions over the course of the semester is consistent with the answer to RQ1 being that the emphasis on calculus in this course may be helping students with their ability with ``applied'' calculus.  Furthermore, the differences observed in student proficiency answering ``pure'' calculus and ``applied'' calculus questions shown in Figure \ref{fig:allresults} and Figure \ref{fig:discrimination} is consistent with student difficulty in mathematization within physics and/or mathematics transfer to physics.  In that context, the results presented here are consistent with previous work demonstrating that such transfer is difficult \cite{britton2002students,cui2006assessing,Rebello2007,new2012researching}, while also indicating that student identity (\textit{i.e.}, gender) may influence transfer proficiency.  

In response to RQ2, the data presented here demonstrate that student perception of calculus ability (\textit{i.e.}, calculus self-efficacy) and student proficiency with calculus is inversely correlated for women, but not for men.  This relationship stems, at least in part, from the fact that while men experience very little change in their calculus self-efficacy over the course of the semester (a Z-statistic of 0.39 for the change from survey 1 to survey 3 for students earning an A in the course), women experience a significant decrease in their calculus self-efficacy over the same period (a Z-statistic of -1.72 for the change from survey 1 to survey 3 for students earning an A in the course).  Thus, even though proficiency with the ``applied'' calculus questions increases more for women (Z-statistic of 4.74 for the change from survey 1 to survey 3) than for men (Z-statistic of 1.92 for the change from survey 1 to survey 2), women nevertheless perceive their proficiency to decrease.  Thus, while an emphasis on calculus in this course may help students improve their proficiency with applied calculus, this emphasis does not appear to impact the calculus self-efficacy of men and negatively impacts the calculus self-efficacy of women.

\subsection{Correlation With Physics Self-Efficacy}

Our observation of a gender imbalance in calculus self-efficacy is consistent with the results of previously reported studies of mathematics self-efficacy \cite{hill2010so,reilly2019investigating}.  Furthermore, the disconnect between calculus self-efficacy and calculus proficiency is reminiscent of previous studies showing a gender-dependent disconnect between physics self-efficacy and physics proficiency (using course grade as a proxy for proficiency)\cite{nissen2016gender,marshman2018female,espinosa2019reducing}.  For example, Marshman \textit{et al.} showed that women had lower physics self-efficacy compared to that of similarly performing men \cite{marshman2018female}.  

To explore this further, we compared the changes in physics self-efficacy to changes in calculus self-efficacy.  Physics self-efficacy was assessed using the instrument previously developed by Marshman \textit{et al.} \cite{marshman2018female}.  This multiple choice survey consists of six questions, each of which involved 4-point Likert scales, with higher scores indicating higher levels of self-efficacy.  A single physics self-efficacy score for each student is determined by averaging the scores for the six questions.  Students completed the assessment at the beginning, in the middle, and at the end of the semester concomitant with completing the other surveys described above.

\begin{figure}[h]
\begin{center}
\includegraphics[width=0.8\textwidth]{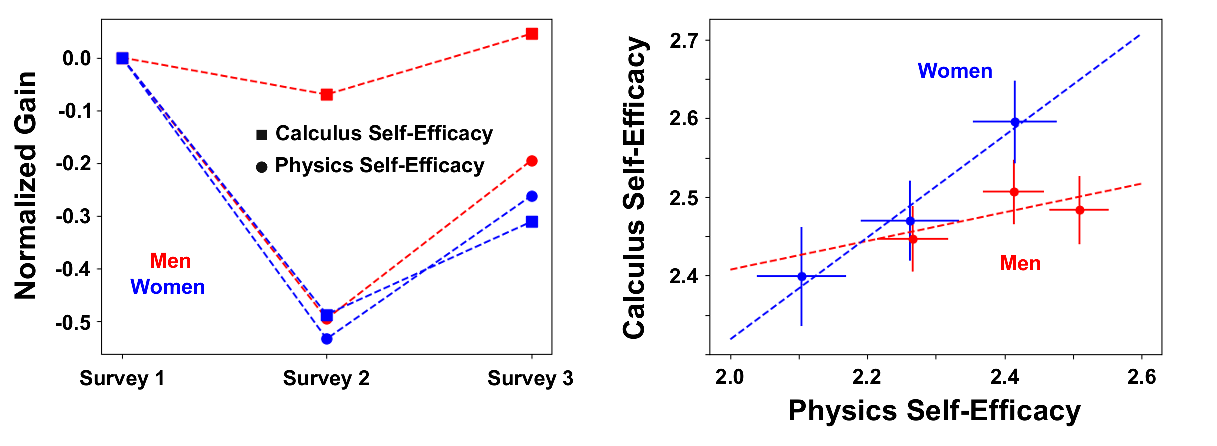}
\end{center}
\caption{Left Panel: The normalized gain in physics self-efficacy (circles) and calculus self-efficacy (squares) for students earning an A in the course.  Right Panel: Correlation between physics self-efficacy and calculus self-efficacy for students earning an A in the course.  The three data points for each cohort correspond to the three deployments of the surveys.}
\label{fig:normalized_gain}
\end{figure}

The normalized gains in the calculus-self efficacy and physics self-efficacy of students earning an A in the course are shown in Figure \ref{fig:normalized_gain}.  For women, the changes in physics self-efficacy and calculus self-efficacy track with one another, possibly suggesting that whatever experiences and/or influences are affected one trait are also affecting the other.  The situation is different for men, however, where the trajectories of changes in physics self-efficacy and calculus self-efficacy do not overlap.  Furthermore, since the changes in physics self-efficacy for men are well aligned with the changes in physics self-efficacy and calculus self-efficacy for women, these data suggest that the trends in calculus self-efficacy for men are the outlier.  One possible explanation for this result is that men react to difficulties using calculus in the course differently from how women do.  For example, men may not associate making mistakes using calculus with a lack of proficiency.  Indeed, their self-efficacy may not decrease if they perceive that other students are also struggling to use calculus since they could associate the making of mistakes as being common in physics courses and thus somehow not a reflection of their intrinsic proficiency with calculus.  That women would react differently to making mistakes using calculus in a physics course could be related to how their perception of success and belonging in that course are influenced by being a member of an underrepresented group in the course and the implicit and/or explicit stereotype threats associated with that status \cite{marchand2013stereotype,maries2018agreeing}.

Our observation that the correlation between physics self-efficacy and calculus self-efficacy is weaker for men than for women (Figure \ref{fig:normalized_gain}), may also suggest that men perceive the role of calculus in the physics course differently than women do.  If so, perhaps such differing views of the role of calculus in physics might be associated not only with the differences in correlation shown in Figure \ref{fig:normalized_gain}, but also with differences in improvement in applied calculus skills (\textit{i.e.}, calculus transfer to physics) during the semester.

\subsection{Limitations}

While the conclusion that these data demonstrate gender-mediated correlations between calculus proficiency and calculus self-efficacy, it is important to note the limitations of this work.  First, the surveys employed in this project were all administered online owing partly to limitations associated with the COVID-19 pandemic.  It is possible that students were not engaged with the surveys, particularly the calculus proficiency survey, as earnestly online as they would have been in-person.  Related to that is the possibility that the preparation obtained by the students in this study in their previously taken calculus courses was affected by those courses being taught online or with a modified curriculum due to pandemic-associated accommodations.  

Second, that fewer than 40\% of the students were answering the ``applied'' calculus questions correctly on the calculus proficiency survey (Figure \ref{fig:allresults} and Figure \ref{fig:discrimination}) may suggest that some of these questions are too difficult or otherwise not able to provide the dynamic range necessary to evaluate student proficiency with using calculus in a physics context.

\subsection{Future Work}

The next step in this project is the further refinement of the calculus proficiency survey.  In addition to potential modification of the questions asked, we would also welcome collaborating with other institutions as that would provide an opportunity to collect data from a larger population of students and potentially a wider distribution of demographics, as well as students completing different physics curricula where expectations for calculus usage may be different.  Such data would also allow us to examine how other dimensions of student identity intersect with calculus proficiency and calculus self-efficacy, as well as the correlation between physics self-efficacy and calculus self-efficacy.

We also intend to coordinate with instructors in other departments to deploy similar surveys of calculus self-efficacy, discipline-specific calculus proficiency, and discipline self-efficacy in their courses.  Data collected from such surveys can help us identify the extend to which the results presented here are specific to introductory physics.  We similarly need to develop and deploy a calculus proficiency survey for General Physics I so that we can identify how early this divergence between calculus self-efficacy and calculus proficiency occurs.

\section{Conclusion}

Since applied mathematics is an important component of all STEM disciplines, there is a clear need to promote student competency with mathematical skills across all STEM curricula.  Introductory physics courses are the ideal venue for improving fluency with applied mathematics as their curricula emphasize quantitative problem solving skills.  Providing opportunities to improve calculus fluency and transfer skills will likely have the largest benefit for students with weaker secondary mathematics preparation \cite{PhysRevPhysEducRes.15.020126}, such as students from traditionally underserved populations \cite{tsoi2015college,atuahene2016mathematics}, who are also at a higher risk of dropping out of STEM degree programs \cite{Moller-Wong1997,Tyson2007,Zhang2004}.  Improving calculus proficiency and transfer ability thus has the potential to improve the retention of these students in these programs and thereby increase the diversity within STEM disciplines and careers.  Similarly, understanding the origins of differences in calculus self-efficacy and the interplay between those differences and variation in calculus proficiency  will assist in the development and assessment of curriculum changes and other interventions targeting improvements in these attributes.  This has the potential to help reduce the imbalances in physics self-efficacy \cite{nissen2016gender,marshman2018female}, leading to an increased participation and retention of traditionally underrepresented students within STEM and, by extension, in the STEM workforce.

\bibliography{library}

\bibliographystyle{abbrv}

\end{document}